\documentclass[aps,prb,twocolumn,superscriptaddress,nofootinbib]{revtex4-2}
\usepackage{amssymb}
\usepackage{amsmath}
\usepackage{mathrsfs}
\usepackage{mathtools}
\mathtoolsset{showonlyrefs=true}
\usepackage{bm}
\usepackage{physics}
\usepackage{enumitem}
\usepackage[svgnames]{xcolor}
\usepackage{float}
\usepackage{tikz}
\usepackage{graphicx}
\usepackage[
    setpagesize=false,
    bookmarks=false,
    colorlinks=true,
    allcolors=magenta
]{hyperref} 
\usepackage[titletoc]{appendix}

\newcommand{\eq}{\mathrm{eq}}

\begin{document}
\title{Impact of time-retarded noise on dynamical decoupling schemes for qubits
}
\author{Kiyoto Nakamura}
\email{kiyoto.nakamura@uni-ulm.de}
\author{Joachim Ankerhold}
\affiliation{Institute for Complex Quantum Systems and IQST, Ulm University, D-89069 Ulm, Germany}

\date{\today}

\begin{abstract}
    One of the simplest and least resource-intensive methods to suppress decoherence for qubit operations, namely, dynamical decoupling (DD), is investigated for a broad range of realistic noise sources with time-retarded feedback.
    By way of example, the Carr--Purcell--Meiboom--Gill (CPMG) sequence is analyzed in a numerically rigorous manner accounting also for correlations between qubits and environments.
    Since experimentally noise sources are characterized through spectral densities, we adopt the spin--boson model as a suitable framework to describe the qubit dynamics under DD for a given spectral density $J(\omega) \propto \omega^s$.
    Motivated by the situation for superconducting qubits, the spectral exponent $s$ is varied from $s=1$ (Ohmic bath) to a substantially small value $0 < s \ll 1$ (deep sub-Ohmic bath), in order to investigate the impact of time-nonlocal back action on DD performances for enhanced coherence times.
    As reference to the DD schemes, dynamics of a single qubit subject to Ramsey sequences without any pules and Hahn echo (HE) sequences are also investigated.
\end{abstract}
\maketitle

\section{Introduction} \label{sec:Introduction}
While remarkable progress in quantum computing has been made in the last decade~\cite{PlaceNATCOMMUN2021,WangNPJQI2022,NegirneacPRL2021,SungPRX2021,KandalaPRL2021,GoogleNATURE2019, GoogleNATURE2023, IBMNATURE2023}, decoherence remains one of the main obstacles to further improvement.
During an idle phase in a quantum algorithm, in which certain qubits are left without any control to wait for the completion of manipulation of other qubits, quantum states of qubits must be conserved for subsequent calculations.
However, those quantum states, especially coherent superpositions, are easily destroyed by environmental noise, which is referred to as decoherence.
Suppression of decoherence hence directly leads to improvements of quantum computation.
In addition to the coherence time during an idle phase (quantum memory) mentioned above, gate operations are affected by decoherence.
During pulse applications, a qubit state starts to deviate from a desired state because of decoherence, which leads to a mismatch of rotation angles and axes.

In order to guide circuit designs and advanced pulse shaping, numerical simulations based on proper noise models are indispensable. The most powerful framework is provided by system--reservoir models (or more specifically, the spin--boson model)~\cite{Weiss2012}, which have begun to be used to characterize decoherence properties and gate fidelities beyond standard weak coupling treatments, see, for example, Refs.~\cite{TuorilaPRR2019,BabuNPJQI2021,GulasciPRB2022,BabuPRR2023,PapicARXIV2023,NakamuraPRB2024,NakamuraPRR2024}, which is the typical situation for solid-state realizations such as superconducting platforms.
The model is advantageous because information about environmental noise is completely encoded in the spectral noise power, which can be experimentally obtained, for example, via noise spectroscopy~\cite{BylanderNP2011,AlmogJPB2011,AlvarezPRL2011} and characterization of quasiparticle noise in superconducting qubits~\cite{RisteNATCOMMUN2013,CardaniNATCOMMUN2021,PanNATCOMMUN2022}.

To further improve the capability of quantum computations, characterization of noise sources can only be the first step:
Quantum control techniques based on this knowledge must be implemented, and eventually quantum error corrections for multi-qubit architectures using syndrome qubits are the ultimate goal.
\textcolor{black}{To remove undesired noise effects during gate operations, pulse shaping techniques based on optimal control theories~\cite{WilhelmARXIV2020} have been proposed.}
A numerical study~\cite{SchmidtPRL2011} for system--reservoir models in which the reservoir dynamics is treated in a numerically rigorous manner has already been conducted, disclosing non-Markovian effects beyond the rotating wave approximation.
\textcolor{black}{Quantum error correction techniques have been extensively studied to correct errors caused by various noise channels~\cite{DevittRPP2013,Lidar2013}, basically addressing bit-flip errors under the assumption of time-local noise.
 The identification of relevant noise channels is thus one of the crucial factors of the quantum error correction.}

\textcolor{black}{In addition to the above methods, noise suppression techniques are also important, especially for noisy quantum devices available nowadays.}
For the improvement of coherence times, the dynamical decoupling (DD) scheme~\cite{ViolaPRA1998} is one of the most widely used techniques fighting against phase errors owing to its simplicity of the implementation.
Originating from techniques for nuclear magnetic resonance (NMR)~\cite{CarrPurcellPhysRev1954,MeiboomGillRST1958}, in which the Hahn echo (HE) sequence~\cite{HahnPR1950} is repeated, a wide variety of DD schemes have been theoretically proposed.
Examples include schemes to suppress noise caused by arbitrary couplings to an environment~\cite{MaudsleyJMR1986,GenovPRL2017}, higher-order contributions of noise in terms of pulse intervals~\cite{KhodjastehPRL2005,UhrigPRL2007}, and errors arising during applications of pulses with finite widths~\cite{ViolaPRL2003}.
Their performances have been tested experimentally~\cite{SouzaPRL2011} in comparison with numerical predictions~\cite{CywinskiPRB2008,QiPRA2023} for different schemes.
In some experiments, the dependence of the performance on different quantum devices has been investigated~\cite{PokharelPRL2018,EzzellPRApplied2023}.
DD schemes with optimization processes for pulse intervals and rotation axes have also been proposed~\cite{BiercukNATURE2009,UysPRL2009,RahmanPRAppl2024}.
\textcolor{black}{Further, optimal control techniques have already been applied to pulse shaping for DD schemes~\cite{YangPRAppl2022}.
It is worth noting that the possibility of improvement of quantum error correction techniques by utilizing DD pulse sequences has been reported in previous studies~\cite{NgPRA2011,Paz-SilvaSCIREP2013}.}

\textcolor{black}{In contrast to this impressive body of literature, DD schemes in presence of experimentally realistic noise sources, i.e., noise with time-retarded feedback as it typically appears in superconducting devices, have not been analyzed in an elaborate manner.}
This seems to be due to the fact that the aim of DD schemes is to universally suppress the undesired time evolution of the system induced by interactions between the qubit system and its environment without any knowledge of coupling forms.
A few examples for specific open systems include studies in which the standard Ohmic~\cite{ViolaPRA1998,UhrigPRL2007} or $1/f$ noise~\cite{CywinskiPRB2008,ShiokawaPRA2004} are considered.
\textcolor{black}{However, a comparison between different types of noise sources including the omnipresent $1/f^{\varepsilon}$-type noise ($\varepsilon \neq 1$) as a major noise source for, for example transmon qubits~\cite{BylanderNP2011,GlazmanSPPLN2021}, has not been presented yet in a comprehensive manner.}

Considering that the DD performance actually depends on the qubit device~\cite{PokharelPRL2018,EzzellPRApplied2023}, an analysis across a broad range of spectral noise powers is essential.
\textcolor{black}{In addition, in most of the previous studies it is assumed  that the time interval between two subsequent pulses $\Delta t$ is sufficiently short compared to typical timescales of the environment $\tau_c$ (short pulses with negligible width, $t_p = 0$, are considered in those studies).
Here, we refer to this assumption $\Delta t \ll \tau_c$ as the short-time assumption, with $\tau_c=1/\omega_c$ with $\omega_c$ being a characteristic frequency of the environment.
However, this is in contrast to typical experimental situations, where the lengths of idle times between subsequent pulses are much longer.}

Another key assumption imposed in previous studies is factorized initial states between the system and reservoir:
\textcolor{black}{This assumption is closely related to the assumption of time-local noise, as both are valid only in the limit of weak system--reservoir coupling at elevated temperature. To arrive at high-precision predictions of the performance of high-fidelity qubit systems, both should be avoided.}

Two questions thus arise: A major one, namely: (Q1) What is the performance of DD schemes beyond the short-time assumption for various noise sources?, and the
implicit subquestion: (Q2) What is the impact of initial correlations between qubits and environments? 
In this paper we provide such an analysis that is expected to give insight for current noisy intermediate-scale quantum (NISQ) devices.
\textcolor{black}{While we specifically refer to superconducting platforms, conceptually, our methodology and the findings apply to  other qubit modalities as well.}

For this purpose, we employ a treatment that does not suffer from any assumption about weak coupling and timescale separation (Born--Markov approximation). More specifically and consistent with previous studies for gate performances~\cite{NakamuraPRR2024}, we provide predictions for performances of Ramsey experiments without pulses, of HE sequences, and of DD schemes.
\textcolor{black}{To focus on the impact of quantum noise, we restrict ourselves to one of the simplest sequences, that is, the Carr--Purcell--Meiboom--Gill (CPMG) sequence~\cite{CarrPurcellPhysRev1954,MeiboomGillRST1958} with impulsive pulses (bang-bang control, $t_p = 0$) for the pure dephasing case.
Effects originating from the finite width of the pulses $t_p \neq 0$ might play an important role when the pulse duration is comparable to the idle time but will be disregarded in this study, i.e., we assume $t_p\ll \Delta t\simeq \tau_c$.}
For these schemes, a broad class of thermal reservoirs relevant for superconducting qubits ranging from reservoirs with Ohmic to those with deep sub-Ohmic characteristics is adopted.
For DD protocols, parameters such as the idle times between two consecutive pulses are tuned over a wide interval from being much shorter to being much longer compared to qubit timescales, in order to guide control schemes for practical NISQ devices.
To better understand the global characteristics of the DD performance, which is defined by the coherence time, we also look into details of the local dynamics of the qubit system. 

This paper is organized as follows.
In Sec.~\ref{sec:model}, we briefly introduce the spin--boson model and its main ingredients relevant for this study.
Since we only consider pure dephasing, analytical expressions for the dynamics of a single qubit are derived.
Section~\ref{sec:results} is devoted to the discussion of the results.
Before we investigate the DD performance, we address the Ramsey experiments without DD pulses and simpler HE sequences as reference.
In Sec.~\ref{sec:RamseyPure}, the contribution of static correlations between the system and the reservoir at an initial time as well as the profile of the spectral noise power to qubit dynamics during the Ramsey sequences is explored.
HE experiments with various spectral noise powers and pulse intervals are investigated in Sec.~\ref{sec:HE}.
Finally, the DD performance for a broad class of  reservoirs is presented in Sec.~\ref{sec:DD}.
\textcolor{black}{We make concluding remarks in Sec.~\ref{sec:conclusion} and discuss how the methodology employed here can be further extended to lift the above assumptions and to provide precise predictions for realistic experimental settings.}

\section{Model and method} \label{sec:model}
In this paper, we consider a single qubit affected by quantum Gaussian noise.
The total Hamiltonian reads
\begin{align}
    \hat{H}_\mathrm{tot} = & \frac{\hbar\omega_q}{2} \hat{\sigma}_z
    - \hbar \hat{\sigma}_z \hat{X} + \hat{H}_R
    = \hat{H}_S - \hat{V}\hat{X} + \hat{H}_R\, .
    \label{eq:H_tot}
\end{align}
Here, the single qubit has been approximated as a two level system, and $\hat{\sigma}_\alpha$ ($\alpha \in \{x, y, z\}$) is the Pauli matrix.
The noise effects are taken into account as the dynamics of a reservoir consisting of an infinite number of harmonic oscillators.
The qubit frequency is given by $\omega_q$ in Eq.~\eqref{eq:H_tot}, and $\hat{H}_R$ is the Hamiltonian of the reservoir.

The coupling between the qubit system and reservoir is in the bilinear form, $\hat{V}\hat{X}$, where $\hat{V} = \hbar \hat{\sigma}_z$ is the system part and $\hat{X}$ is the reservoir part.
Note that we adopt the pure-dephasing coupling for the simplicity.
Considering that the population-relaxation time is much longer than the coherence time in a lot of qubit devices, the above form for $\hat{V}$ might be plausible approximation.
The reservoir is characterized by the autocorrelation function $C(t) = \mathrm{tr}\{\hat{X}(t)\hat{X}(0) \hat{\rho}_{R, \eq}\}$, where $\hat{\rho}_{R, \eq}$ is the Boltzmann distribution of the reservoir, $\hat{\rho}_{R, \eq} = e^{-\beta \hat{H}_R} / \mathrm{tr}\{e^{-\beta \hat{H}_R}\}$ ($\beta = 1 / k_\mathrm{B} T$ is the inverse temperature with the Boltzmann constant $k_\mathrm{B}$).
More specifically, we introduce the spectral density
\begin{align}
    J(\omega) = & \frac{1-e^{-\beta \hbar \omega}}{\hbar} S_\beta (\omega)
    = \frac{1-e^{-\beta \hbar \omega}}{2 \pi \hbar}
    \int_{-\infty}^{\infty} dt C(t) e^{i\omega t}\, ,
\end{align}
and the reservoir is characterized through the parameters of $J(\omega)$. Here, $S_\beta(\omega)$ is the spectral noise power.

To express the sequence of the HE and DD simply, we define the following superoperators:
the rotation superoperator of the system on the Bloch sphere,
\begin{align}
    \mathcal{R}_x \hat{\rho}_\mathrm{tot}(t) = 
    \hat{R}_x(\pi) \rho_\mathrm{tot}(t) \hat{R}_x(-\pi) \, ,
    \label{eq:piPulse}
\end{align}
and the time-evolution superoperator according to Eq.~\eqref{eq:H_tot},
\begin{align}
    \mathcal{U}_\mathrm{i}(\Delta t) \hat{\rho}_\mathrm{tot}(t) = 
    e^{-i\hat{H}_\mathrm{tot} \Delta t / \hbar} \hat{\rho}_\mathrm{tot}(t)
    e^{i\hat{H}_\mathrm{tot} \Delta t / \hbar}\, .
\end{align}
The former corresponds to the rotation about the $x$ axis by the angle $\pi$ with an impulsive pulse, while the latter is the idling without pulses.
\textcolor{black}{In all the following simulations, we only consider impulsive pulses (instantaneous rotations) to eliminate errors during pulse applications and focus on the effects of the free evolution $\mathcal{U}_\mathrm{i}(\Delta t)$.}
The operator $\hat{R}_\alpha(\theta) = \exp[-i \theta \hat{\sigma}_\alpha/2]$ in Eq.~\eqref{eq:piPulse} is the rotation operator about an $\alpha$ axis by an angle $\theta$ and is applied to the single-qubit subsystem.
\textcolor{black}{Note that the form of the operator for the impulsive $\pi$ pulse is modified depending on whether the laboratory or rotating frame is considered, which makes significant differences between theoretical predictions and experimental results~\cite{TripathiPRApplied2022}.
For example, when one considers the application of a continuous microwave with the frequency $\omega_\mathrm{ex}$ to a transmon qubit as in Ref.~\cite{BlaisRMP2021}, the corresponding impulsive $\pi$ pulse is $\hat{R}_x(\pi)$ in the \emph{rotating} frame.
Since Eq.~\eqref{eq:H_tot} is in the laboratory frame, the rotation operator $\hat{R}_x(\pi)$ in Eq.~\eqref{eq:piPulse} must be modified to $\hat{R}_z(\omega_\mathrm{ex} t) \hat{R}_x(\pi) \hat{R}_z(-\omega_\mathrm{ex} t)$~\cite{NakamuraPRR2024} in this situation.
Following most of the DD schemes devised in previous theoretical studies, however, we here assume that the operator is expressed as $\hat{R}_x(\pi)$ in the \emph{laboratory} frame, obtaining Eq.~\eqref{eq:piPulse}.}

Since the system Hamiltonian and the system part of the interaction commute, $[\hat{H}_S, \hat{V}] = 0$, the dynamics of the system for the Ramsey, HE, and DD experiments are expressed in an analytical form, see Sec.~\ref{sec:results} and Appendix~\ref{sec:appEcho} for more details.
\textcolor{black}{We emphasize that because of the generality of Eq.~\eqref{eq:H_tot}, the analytical expressions in Sec.~\ref{sec:results} and Appendix~\ref{sec:appEcho} can be applied to \emph{general two-level systems subject to pure-dephasing noise}.}

Now, we move on by specifying a form of the spectral density and parameter values.
The spectral density is given by
\begin{align}
    J_s(\omega) = \mathrm{sgn}(\omega)
    \frac{\kappa \omega_\mathrm{ph}^{1-s} |\omega|^{s}}
    {\bigl[1+(\omega/\omega_c)^2\bigr]^2}\, ,
\end{align}
where $\omega_c$ and $\kappa$ are the cutoff frequency and the coupling strength between the system and reservoir, respectively.
The spectral exponent $s$ determines the behavior of $J_s(\omega)$, and the quantity $\omega_\mathrm{ph}$ has been introduced to fix the unit of $\kappa$ irrespective of $s$.

Depending on the spectral exponent $s$, the spectral density is classified into the following three types:
Ohmic ($s=1$), super-Ohmic ($s > 1$), and sub-Ohmic ($0 < s < 1$) types.
The spectral noise power for the Ohmic bath $S_{\beta, s=1}(\omega)$ corresponds to the white noise in the low frequency region, $S_{\beta, s=1}(\omega = 0) = \kappa k_\mathrm{B} T$, while the sub-Ohmic baths represent $1/f^\varepsilon$~noise, $S_{\beta, s < 1}(\omega \to 0) \simeq \kappa k_\mathrm{B}T (\omega_\mathrm{ph}/\omega)^{1-s}$.
The most common class of noise for many qubit modalities is environmental fluctuations modeled as Ohmic reservoirs~\cite{Weiss2012,Barone1982,WendinARXIV2005}, while specifically for superconducting qubits also $1/f^\varepsilon$~noise is found~\cite{IthierPRB2005,BylanderNP2011} as a result of two-level fluctuators~\cite{MachlupJAP1954,Weiss2012,PaladinoRMP2014,MullerRPP2019} and quasiparticles' dynamics~\cite{GlazmanSPPLN2021}.
To cover a broad range of spectral distributions, we choose the values $1$, $1/2$, $1/4$, $1/8$, and $1/14$ for the exponent $s$ in this study.
\textcolor{black}{Further, we set $\omega_q$ as the unit of  frequency and use dimensionless parameter values for the reservoirs as $\beta \hbar \omega_q = 5$, $\omega_c = 50 \omega_q$, $\omega_\mathrm{ph} = \omega_{q}$, and $2\pi\hbar\kappa = 0.04$.
We use these values here by way of example, and they can be easily modified in the analytical expressions given in  Sec.~\ref{sec:results}.}


For the numerical simulations, it is convenient to expand the autocorrelation function of the reservoir in a mathematically consistent way as follows:
\begin{align}
    C(t) = \sum_{k=1}^{K} d_k e^{-i\omega_k t-\gamma_k t} \qquad (t > 0)\, ,
    \label{eq:CF}
\end{align}
which was introduced to the free-pole hierarchical equations of motion (FP-HEOM) method~\cite{XuPRL2022}.
To obtain the sets $\{d_k\}$, $\{\omega_k\}$, and $\{\gamma_k\}$ efficiently for a given accuracy, 
the barycentric representation of the spectral noise power $S_\beta(\omega)$ is exploited.
\textcolor{black}{Other representations of $C(t)$, for example, those reported in a previous study~\cite{TakahashiJCP2024}, are also, of course, applicable to this study.}



\section{Results}\label{sec:results}
In this section, we present results of the Ramsey experiments and investigate whether and to what extent HE and DD improve the coherence time in presence of time-retarded noise feedback (Q1). 

\subsection{Ramsey experiments}\label{sec:RamseyPure}
Before we turn to the analysis of the numerical results, we make a brief remark on the different types of the initial states:
We consider factorized and correlated initial states in this study.
The factorized initial state is defined as 
\begin{align}
    \hat{\rho}^\mathrm{f}_\mathrm{tot}(0)  = & \hat{R}_y\Bigl(-\frac{\pi}{2}\Bigr)
    \ketbra{0}{0} \hat{R}_y \Bigl(\frac{\pi}{2}\Bigr) \otimes \hat{\rho}_{R, \eq} \\
    = & \frac{\bigl(\ket{0}+ \ket{1}\bigr)
    \bigl(\bra{0} + \bra{1}\bigr)}{2} \otimes \hat{\rho}_{R, \eq}\, ,
    \label{eq:initEchoF}
\end{align}
and the correlated one as 
\begin{align}
    \hat{\rho}^\mathrm{c}_\mathrm{tot}(0) = \hat{R}_y\Bigl(-\frac{\pi}{2}\Bigr)
    \frac{e^{-\beta \hat{H}_\mathrm{tot}}}{{\mathrm{tr}\{e^{-\beta\hat{H}_\mathrm{tot}}\}}}
    \hat{R}_y \Bigl(\frac{\pi}{2}\Bigr)\, .
    \label{eq:initEchoC}
\end{align}
Here, the state $\ket{0}$ is the ground state of the qubit system (correspondingly, $\ket{1}$ indicates the first-excited state).
In the latter preparation, all correlations caused by the total equilibrium state $e^{-\beta \hat{H}_\mathrm{tot}}$ are taken into account.
The former initial state is based on the assumption that during the preparation of the system into the ground state $\ketbra{0}{0}$, equilibrium correlations caused by $e^{-\beta \hat{H}_\mathrm{tot}}$ are destroyed.

\begin{figure}
    \centering
    \includegraphics[width=\linewidth]{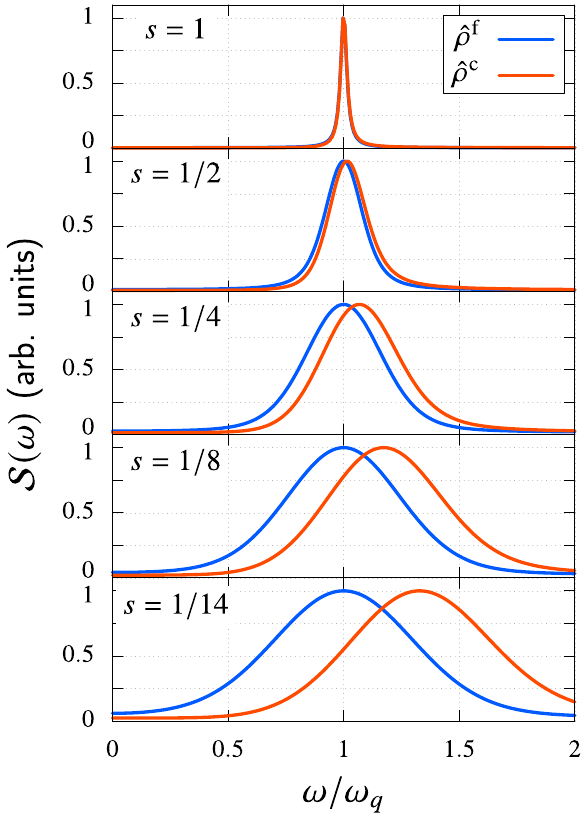}
    \caption{Fourier transform of $\ev*{\hat{\sigma}_x(t)}$ obtained with the Ramsey experiments, $\mathcal{S}(\omega)$, with arbitrary (arb.) units.
    Each curve is normalized such that the maximum value is $1$.
    The results with the factorized ($\hat{\rho}^\mathrm{f}$, blue curve) and correlated ($\hat{\rho}^\mathrm{c}$, red curve) initial states are depicted.
    \label{fig:RamseyFFTPure}}
\end{figure}

We start with the qubit spectroscopy assuming these different initial states in  Fig.~\ref{fig:RamseyFFTPure}.
It depicts the Fourier transform of $\ev*{\hat{\sigma}_x(t)} = \mathrm{tr}\{\hat{\sigma}_x \mathcal{U}_\mathrm{i}(t) \hat{\rho}_\mathrm{tot}(0)\}$,
\begin{align}
    \mathcal{S}(\omega) = \frac{1}{2} \mathrm{Re}\biggl\{\int_{0}^{\infty} dt \ev*{\hat{\sigma}_x(t)}\, e^{-i \omega t}\biggr\}\, ,
    \label{eq:FFT}
\end{align}
where $\hat{\rho}_\mathrm{tot}(0) = \hat{\rho}^\mathrm{f}_\mathrm{tot}(0)$ (the blue curve) and $\hat{\rho}^\mathrm{c}_\mathrm{tot}(0)$ (the red curve).
There, one finds that the width of the profile is broader for the smaller spectral exponent $s$.
The peak of $\mathcal{S}(\omega)$ for the factorized initial state is strictly at the value $\omega_q$, while in the case with the correlated initial state, the peak is shifted to a larger frequency.
In the following, we analyze these properties through the analytical expression of the reduced density operator (RDO) of the system.

Defining the off-diagonal element of the system RDO as $\rho_{eg}(t) = \mel{1}{\hat{\rho}_S(t)}{0} = \mel{1}{\mathrm{tr}_R\{\hat{\rho}_\mathrm{tot}(t)\}}{0}$, we evaluate the expectation value $\ev*{\hat{\sigma}_x(t)}$ through the equation $\ev*{\hat{\sigma}_x(t)} = 2 \mathrm{Re}\{\rho_{eg}(t)\}$.
Here, $\mathrm{tr}_R\{\bullet\}$ indicates the partial trace over the reservoir degrees of freedom.
The dynamics of the off-diagonal element are derived as follows:
for the factorized initial state,
\begin{align}
    \rho_{eg}(t)
    = \frac{1}{2}\exp\Biggl[-i\omega_q t-4\int_{0}^{t}dt'\int_{0}^{t'}dt'' C'(t'-t'') \Biggr]
    \, , \quad
    \label{eq:offDiagF}
\end{align}
and for the correlated initial state,
\begin{align}
    \rho_{eg}(t) = \rho_{+}(t) - \rho_{-}(t)\, ,
    \label{eq:offDiagC}
\end{align}
where the contributions of the ground and excited states, $\rho_{+}(t)$ and $\rho_{-}(t)$, are given by
\begin{align}
    \rho_{\pm}(t)
    = \frac{e^{\pm\beta\hbar\omega_q/2}}{2Z} \exp\Biggl[
    \begin{aligned}[t]
        & -i\omega_q t \mp 4i \int_{0}^{t}dt' \bar{L}(t') \\
        & -4\int_{0}^{t}dt'\int_{0}^{t'}dt'' C'(t'-t'') \Biggr]\, .
    \end{aligned}
    \\
    \label{eq:FIDC}
\end{align}
Note that the function $\bar{L}(t) = -\int_{t}^{\infty} dt' C''(t')$ is proportional to the relaxation function of the reservoir~\cite{Kubo1985}.
Here, the real and imaginary parts of the autocorrelation function are expressed through the relation $C(t) = C'(t) + i C''(t)$.
The quantity $Z$ is the partition function of the bare system, which is described as $Z = \mathrm{tr}\{e^{-\beta\hat{H}_S}\} = 2 \cosh(\beta \hbar \omega_q/2)$.

First, we investigate the width of the profile.
The term $\int dt' \int dt'' C'(t'-t'')$ in Eqs.~\eqref{eq:offDiagF} and \eqref{eq:FIDC} contributes to the dynamics as decoherence, and we found that this integral is larger for smaller $s$ in our study.
This results in the broader peak for the smaller spectral exponent.
Note that within the Born--Markov approximation, in the Ohmic and super-Ohmic cases, the decay rate caused by decoherence, $1/T_\mathrm{R}$, is proportional to $S_\beta(0)$ (see Appendix~\ref{sec:appEcho} and Refs.~\cite{Weiss2012, Breuer2002}).
In the sub-Ohmic case, however, the spectral noise power diverges at $\omega = 0$, and this simple argument stemming from a perturbative treatment cannot be applied.
In fact, the stronger divergence of $S_\beta(\omega)$ around $\omega\to 0$ for smaller exponents directly correlates with the tendency of $T_\mathrm{R}$ (cf. Table~\ref{tbl:timeConst} below for the time constant with the factorized initial state).
Note that the width is the same for both factorized and correlated initial states.

Next, we focus on the frequency shift.
For the factorized initial state, the reservoir does not change the Larmor frequency, as indicated by Eq.~\eqref{eq:offDiagF}, and the peak of each blue curve is found at $\omega = \omega_q$ in Fig.~\ref{fig:RamseyFFTPure}.
By contrast, the effective Larmor frequency is modified with the term $i \int dt' \bar{L}(t')$ in Eq.~\eqref{eq:FIDC} in the case with the correlated initial state.
Because we consider the low temperature $\beta \hbar \omega_q = 5$, the contribution of $\rho_{-}(t)$ to $\rho_{eg}(t)$ in Eq.~\eqref{eq:offDiagC} is small.
The frequency of $\rho_{+}(t)$ at the time $t$ is given by $\omega_q + 4 \int_{0}^{t}dt' \bar{L}(t')$ in Eq.~\eqref{eq:FIDC}, and we found that the function $\bar{L}(t)$ always takes non-negative values in our simulations.
For this reason, the peak for the correlated initial state is shifted to a larger frequency, which is indicated by the red curves in Fig.~\ref{fig:RamseyFFTPure}.
The function $\bar{L}(t)$ decays slower as the spectral exponent $s$ decreases, and hence the integral of $\bar{L}(t)$ takes larger values.
This results in a remarkable shift of the red curve in Fig.~\ref{fig:RamseyFFTPure} for smaller exponents.
\textcolor{black}{The shift in this domain is on the order of the qubit frequency, which has not been seen in actual experiments, for example with superconducting platforms. We thus conclude that deep sub-Ohmic reservoirs do not dominantly couple in this situation.}

Note that the Lamb shift found in the conventional Born--Markov approximation approach~\cite{RedfieldIJRD1957,LindbladCMP1976,Breuer2002} is independent of the above frequency shift:
The Lamb shift results from correlations of the system and reservoir in the real-time domain, while the above frequency results from correlations between the total equilibrium state ($e^{-\beta \hat{H}_\mathrm{tot}}$) and the real-time evolution [$\mathcal{U}_\mathrm{i}(t)$].
\textcolor{black}{Since within the Born--Markov approximation the total system is always factorized into the form $\hat{\rho}_S(t) \otimes \hat{\rho}_{R, \eq}$, the frequency shift caused by the initial correlation between the system and reservoir cannot be seen.}

\textcolor{black}{Finally, we briefly remark on the dependence of the profile in Fig.~\ref{fig:RamseyFFTPure} on the coupling strength between the system and reservoir, $\kappa$, and the cutoff frequency, $\omega_c$.
As indicated by Eqs.~\eqref{eq:offDiagF} and \eqref{eq:FIDC}, broader line widths and larger frequency shifts are observed for larger coupling strengths.
This indicates that the magnitude of the frequency shift highly depends on both the spectral exponent $s$ and the coupling strength $\kappa$, so that  $s$ cannot be extracted solely by determining the frequency shift.
The cutoff frequency $\omega_c$ hardly changes the profile, which contrasts with the coupling strength $\kappa$: 
The dynamics of $\rho_{eg}(t)$ in the time domain vary according to the cutoff frequency $\omega_c$ during the time period $t \lesssim 1/\omega_c$, but since the long-time behavior is determined by the other parameters, such as the temperature $1/\beta$, coupling strength $\kappa$, and spectral exponent $s$~\cite{Breuer2002,Weiss2012}, the profiles in the frequency domain are almost the same irrespective of $\omega_c$.
Note that however, the cutoff frequency $\omega_c$ plays a crucial role in the HE and DD dynamics, as discussed below.}

In the following numerical experiments, we evaluate the absolute value of the off-diagonal element $|\rho_{eg}(t)|$ to explore the impact of decoherence (dephasing).
Due to the negligible contribution of $\rho_{-}(t)$, the absolute value is well approximated as $|\rho_{eg}(t)| \simeq |\rho_{+}(t)|$ at the low temperature.
\textcolor{black}{Since it turns out that for the given set of parameters differences in the normalized dynamics $|\rho_{eg}(t)|/|\rho_{eg}(0)|$ between the factorized and correlated initial states are negligibly small (the largest absolute difference of $|\rho_{eg}(t)|/|\rho_{eg}(0)|$ between those initial states is on the order of $10^{-3}$), we only discuss the former case in detail in the following study.}

\begin{table}
    \caption{Time constants of the decay of the off-diagonal element $|\rho_{eg}(t)|$ obtained from the Ramsey and echo experiments. The quantity $T_\mathrm{R}$ is the time constant of the Ramsey experiments, and $T_\mathrm{E}$ is the time constant of the echo experiments.
    \label{tbl:timeConst}}
    \begin{ruledtabular}
    \begin{tabular}{cccccc}
        $s$ & $1$ & $1/2$ & $1/4$ & $1/8$ & $1/14$  \\ 
        \hline 
        $\omega_q T_\mathrm{R}$ & $62.5$ & $16.9$ & $7.82$ & $5.39$ & $4.46$ \\
        $\omega_q T_\mathrm{E}$ & $62.5$ & $27.7$ & $19.6$ & $17.2$ & $16.5$ \\
    \end{tabular}        
    \end{ruledtabular}
\end{table}

\begin{figure*}
    \centering
    \includegraphics[width=1\linewidth]{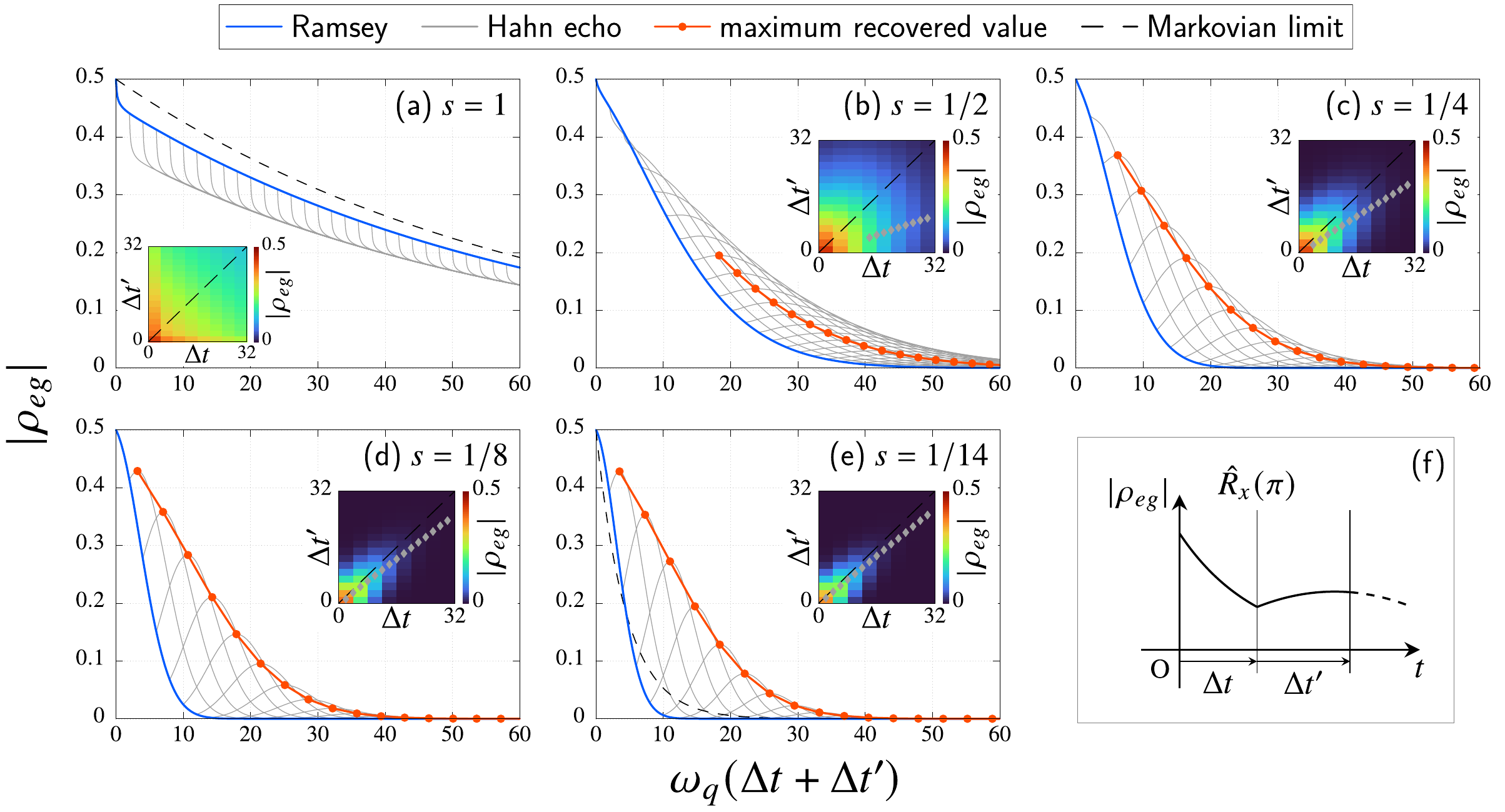}
    \caption{\textcolor{black}{\textcolor{black}{(a)--(e)} Evolution of $|\rho_{eg}(\Delta t + \Delta t')|$ during the echo experiments for various fixed idle times $\Delta t$ and varying $\Delta t'$ (gray curves). Decay dynamics without any pulses (Ramsey experiments) are depicted as blue curves, and the points at which the gray curves begin to deviate from these curves correspond to the pulse-application time $\Delta t$. The red filled circles indicate a local maximum of $|\rho_{eg}(t)|$ after the pulse application with a given $\Delta t$. In the case \textcolor{black}{(a)} $s = 1$ (Ohmic reservoir), no recovery of the coherence occurs and filled circles are not depicted. Black dashed curves in the panels for \textcolor{black}{(a)} $s = 1$ and \textcolor{black}{(e)} $1/14$ refer to the dynamics obtained in the Markovian limit (see Appendix~\ref{sec:appEcho} for details). Inset: Heat maps of $|\rho_{eg}(\Delta t, \Delta t')|$ as a function of both $\Delta t$ and $\Delta t'$. The diagonal dashed line, $\Delta t' = \Delta t$, is depicted as a guide to the eye. The gray diamonds correspond to the red filled circles in the main plot. \textcolor{black}{(f)} Pulse scheme for the Echo experiment.}
    \label{fig:echo3d}}
\end{figure*}

\subsection{Hahn echo (HE)} \label{sec:HE}
In this section, we discuss the numerical results of the HE.
The sequence of the HE is given by
\begin{align}
    \mathcal{U}_\mathrm{i}(\Delta t') \mathcal{R}_x
    \mathcal{U}_\mathrm{i}(\Delta t)\, ,
    \label{eq:seqEcho}
\end{align}
and the schematic of this sequence is displayed in \textcolor{black}{Fig.~\ref{fig:echo3d}(f)}.
In the same way as the Ramsey experiment, the dynamics of $\rho_{eg}(t)$ is analytically evaluated, for details see Appendix~\ref{sec:appEcho}.

\textcolor{black}{\textcolor{black}{Figures~\ref{fig:echo3d}(a)--\ref{fig:echo3d}(e)} displays the dynamics of the off-diagonal element of the RDO, $|\rho_{eg}(t_\mathrm{e})|$, starting at $|\rho_{eg}(t_\mathrm{e}=0)|=1/2$, where the time argument $t_\mathrm{e}$ is given by $t_\mathrm{e} = \Delta t + \Delta t'$ [end of a single HE sequence, cf.\ \textcolor{black}{Fig.~\ref{fig:echo3d}(f)}].
The blue curves correspond to the Ramsey experiments without the $\hat{R}_x(\pi)$ pulse, and the gray curves correspond to the dynamics after the pulse application ($\Delta t \leq t \leq t_\mathrm{e}$): each gray curve corresponds to a different waiting time $\Delta t$.
}

A general result of our simulations is that a recovery of $|\rho_{eg}(t)|$ is observed for sub-Ohmic reservoirs only, while the application of a pulse is always detrimental for Ohmic ones:
\textcolor{black}{Step-like decay right after the pulse application is observed in the Ohmic case.}
In order to understand this, one has to recall that the echo technique was originally developed to reduce the effects of the ``inhomogeneous broadening'' in NMR experiments.
This broadening results from the inhomogeneity of \emph{static} magnetic fields, which, within the framework of the system--reservoir model, corresponds to a time-independent two-time correlation function of the reservoir in the classical limit~\cite{TanimuraJPSJ06}.
Now, for sub-Ohmic spectral densities the two-time correlator decays slower for smaller spectral exponents $s$, which is to say that the reservoir tends to become gradually more sluggish.
Accordingly, in the Ohmic case ($s=1$), the relatively fast decay of the two-time correlation function is dominated by the large portion of higher-frequency modes in the region $\omega_q \ll \omega \lesssim \omega_c$, leading to the fast decay of $|\rho_{eg}(t_\mathrm{e})|$ in the HE experiments.
With the growing portion of low-frequency modes in the spectral noise power, the strongly suppressed dynamics of the reservoir make it possible to mitigate the impact of decoherence by echo sequences. In order for this to be effective, the spectral exponent has to be lower than $s\approx 1/2$ since for this value 
the recovery is not yet observed in the short-time region $\omega_q \Delta t < 14$ \textcolor{black}{[cf.\ the gray curve in \textcolor{black}{Fig.~\ref{fig:echo3d}(b)} for $s = 1/2$ with $\Delta t = 2$, which indicates fast decay right after the pulse application around the time $\Delta t' \simeq 0$].
Note that the performance of HE strongly depends on the cutoff frequency $\omega_c$: We found that for a smaller $\omega_c$, the recovery is observed for $s = 1/2$ even in the short-time region $2 \leq \omega_q \Delta t \leq 14$ (results are not shown).}

To investigate to what extent the echo sequence improves the qubit's coherence time, we depict in \textcolor{black}{Figs.~\ref{fig:echo3d}(b)--\ref{fig:echo3d}(e)} the maximum recovered value of $|\rho_{eg}(t)|$ (red filled circles):
for a fixed $\Delta t$, we detect local maxima with respect to $\Delta t'$, and monitor the value of each local maximum as a function of the total time $\Delta t + \Delta t'$. These results are then compared with those of Ramsey experiments (the blue curves, cf. Sec.~\ref{sec:RamseyPure}).
The time constants extracted from this analysis are listed in Table~\ref{tbl:timeConst}, see also Appendix~\ref{sec:appTimeConst}.
As already mentioned above, in the Ohmic case, the intensity significantly decreases right after the pulse application, where the overall decay time remains almost the same.
By contrast, in the sub-Ohmic cases, the improvement of the time constant is more significant with an increasing ratio $T_\mathrm{E} / T_\mathrm{R}$ for smaller spectral exponents.
This implies that since the time-independent properties of the two-time correlation function are enhanced in the deep sub-Ohmic domain, the echo sequence works in a more efficient way.
\textcolor{black}{When the correlation function tends to be almost time-independent, the coherence is recovered to a maximum at the time $\Delta t' = \Delta t$.
This corresponds to the peak positions aligning along the diagonal when the off-diagonal element is plotted as a function of $\Delta t$ and $\Delta t'$ in the form of $|\rho_{eg}(\Delta t, \Delta t')|$.
The insets of \textcolor{black}{Figs.~\ref{fig:echo3d}(a)--\ref{fig:echo3d}(e)} are the corresponding heat maps, and it is again verified that a deep sub-Ohmic reservoir behaves almost as a static reservoir on relevant timescales.}
Conversely, if we measure the signal with the condition $\Delta t = \Delta t'$, the time constants decrease compared to what is depicted in \textcolor{black}{Figs.~\ref{fig:echo3d}(a)--\ref{fig:echo3d}(e)}.

Note that while the improvement of the time constant from the Ramsey to the HE experiments is the most significant in the case $s=1/14$, the absolute value of the time constant for the Ramsey experiment in this case is the smallest, as discussed in Sec.~\ref{sec:RamseyPure}.

\textcolor{black}{Finally, we compare the above exact results with approximate dynamics.
The black dashed curves in \textcolor{black}{Figs.~\ref{fig:echo3d}(a) and \ref{fig:echo3d}(e)} for $s = 1$ and $1/14$ are the \emph{echo} dynamics obtained with the Markov approximation (for the derivation of the equation of motion, see Appendix~\ref{sec:appEcho}): neither recovery ($s=1/14$) nor immediate decay ($s = 1$) occurs because of the lack of memory effects, and the dynamics coincide with the results of the \emph{Ramsey} experiments obtained with the Markov approximation.
This leads us to the estimation of errors introduced into the Ramsey results by the Markov approximation through the comparison between the blue and black curves in \textcolor{black}{Figs.~\ref{fig:echo3d}(a) and \ref{fig:echo3d}(e)}.
Within the Markov approximation, the decoherence dynamics are always represented by a single exponential function, and hence the fast relaxation process for $s=1$ and Gaussian-like decay for $s=1/14$ cannot be expressed within the approximation.}

\subsection{Dynamical decoupling (DD)}\label{sec:DD}
\begin{figure*}
    \includegraphics[width=\linewidth]{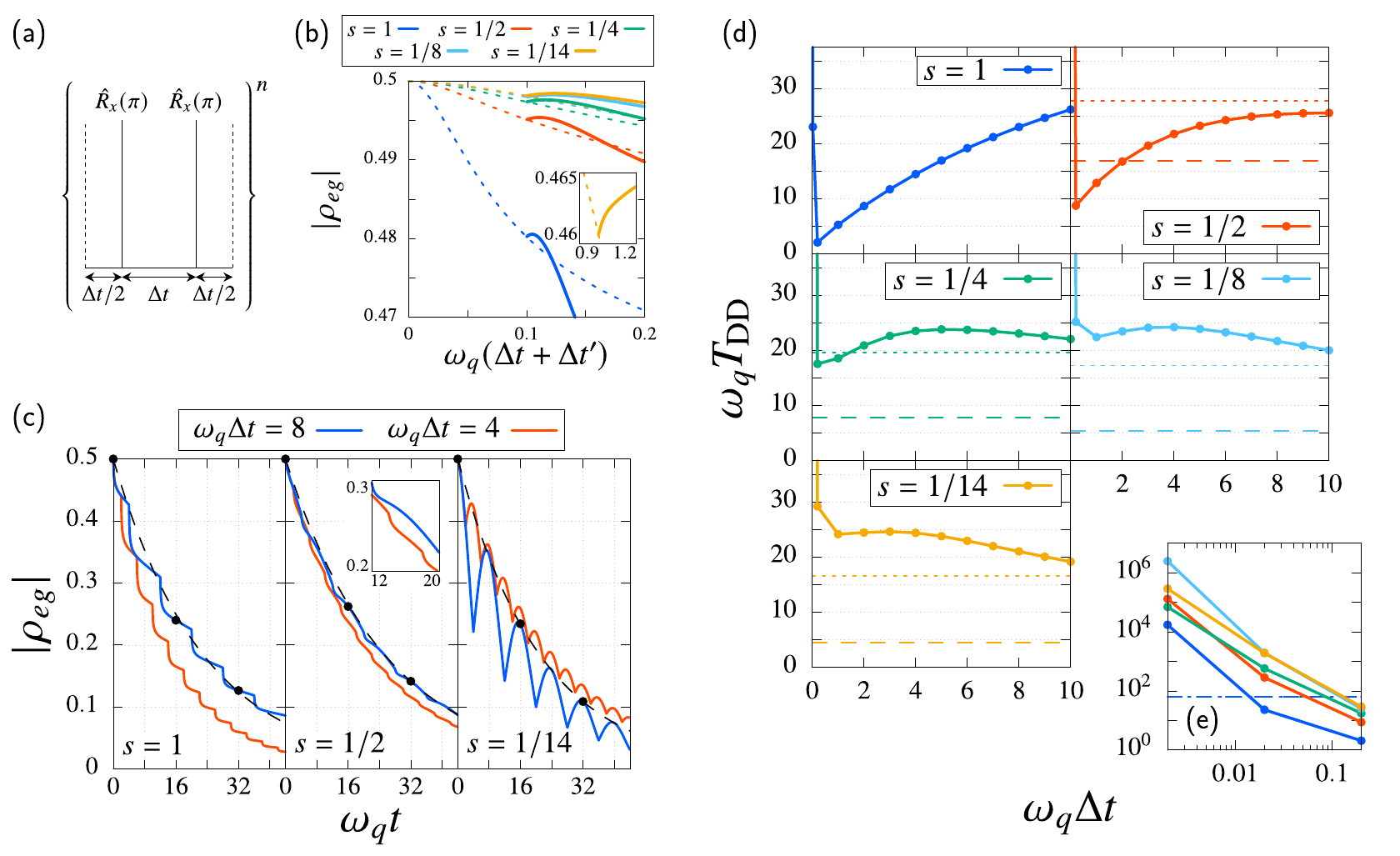}
    \caption{(a) Schematic of the CPMG sequence. (b) Dynamics of $|\rho_{eg}(t)|$ during the Hahn-echo experiments in the short-time region. The duration is given by $\omega_q \Delta t = 0.1$. The dashed curves indicate the dynamics of the Ramsey experiments. In the inset, the same dynamics but with $\omega_q \Delta t = 1$ and $s = 1/14$ are plotted. (c) Dynamics of the off-diagonal element $|\rho_{eg}(t)|$ during the dynamical-decoupling experiments. As representatives, the cases $\omega_q \Delta t = 4$ and $8$ for the spectral exponent $s = 1$, $1/2$, and $1/14$ are depicted. \textcolor{black}{The filled circles indicate the end of each cycle in the case $\omega_q \Delta t = 8$, and the dashed curves are fitted exponentials obtained from the values of the filled circles (see Appendix~\ref{sec:appTimeConst}). Inset for $s = 1/2$: Magnification of the region $12 \leq \omega_q t \leq 20$} (d) Time constant of the decay $T_\mathrm{DD}$ with respect to the duration of the idle phase $\Delta t$. In the sub-Ohmic case, the time constants for the Ramsey and echo experiments, $T_\mathrm{R}$ and $T_\mathrm{E}$, which are listed in Table~\ref{tbl:timeConst}, are also depicted as the dashed and dotted line, respectively. (e) Log--log plot of $T_\mathrm{DD}$ in the small-$\Delta t$ region. The time constants, $T_\mathrm{R}$ and $T_\mathrm{E}$, in the Ohmic case are depicted here as the dashed and dotted blue line, respectively.
    \label{fig:dd}}
\end{figure*}
Next, we study the dynamics during DD experiments with the pulse sequence 
\begin{align}
    \Bigl\{\mathcal{U}_\mathrm{i}(\Delta t /2)
    \mathcal{R}_x \mathcal{U}_\mathrm{i}(\Delta t)
    \mathcal{R}_x \mathcal{U}_\mathrm{i}(\Delta t/2)
    \Bigr\}^n  \quad
    \label{eq:seqDD}
\end{align}
with integer $n>1$. The schematic of this sequence is shown in Fig.~\ref{fig:dd}(a).
Here, we follow previous studies~\cite{ViolaPRA1998,ShiokawaPRA2004,RegoARPC2009}, and a symmetric version of the  Carr--Purcell--Meiboom--Gill (CPMG) sequence is applied.
Notably, while in previous studies~\cite{ViolaPRA1998,ShiokawaPRA2004,RegoARPC2009,RomeroQST2024} only small to moderate durations $\Delta t$ have been considered, here, we explore a wide range of idle durations from small values much below the qubit frequency to relatively large values beyond, i.e., $0.002 \leq \omega_q \Delta t \leq 10$.
We again refer the readers to Appendix~\ref{sec:appEcho} for the analytical expression of the dynamics of $\rho_{eg}(t)$ during the sequence of Eq.~\eqref{eq:seqDD}.

By way of example, the dynamics with $\omega_q \Delta t = 4$ and $\omega_q \Delta t =8$ for spectral exponents $s = 1$, $1/2$, and $1/14$ are depicted in Fig.~\ref{fig:dd}(c). Strikingly, one observes a different behavior for decreasing $s$ when $\Delta t$ grows: in the Ohmic regime a larger $\Delta t$ implies improved coherences while the opposite is true in the deep sub-Ohmic regime. In order to explore this quantitatively, we extract from these time traces time constants $T_\mathrm{DD}$ for the decay of coherences in the presence of DD with varying $\Delta t$ (for details see Appendix~\ref{sec:appTimeConst}) and depict them in Fig.~\ref{fig:dd}(d).
The total time of the time evolution for the evaluation of $T_\mathrm{DD}$ is fixed to $200 / \omega_q$, so that we can obtain the total number of pulses applied during the whole sequence accordingly (from $20$ to $10^5$).

In all cases, irrespective of the spectral exponent $s$, we can distinguish a domain for very short times $\omega_q \Delta t \ll 1$ with a steep drop of $T_\mathrm{DD}$ and a domain beyond with much smoother behavior. In the latter domain the time constant $T_\mathrm{DD}$ increases and, deeper in the sub-Ohmic regime, approaches a local maximum before it tends to decrease again. 
The transition between these two time domains appears as a local minimum, the position of which shifts towards $\omega_q \Delta t\approx 1$ for decreasing spectral exponents.
\textcolor{black}{Note that the time interval $\Delta t$ at which $T_\mathrm{DD}$ takes a local minimum directly depends on the cutoff frequency $\omega_c$: we found local minima at smaller $\Delta t$ for larger $\omega_c$, while at larger $\Delta t$ for smaller $\omega_c$ (results are not shown).}


In order to understand this behavior, we first turn to the very-small-$\Delta t$ region. There, one can consider the simplified pulse sequence of the HE, as discussed above, since the HE sequence also appears in the fundamental pulse sequence for the DD, cf.\ Eqs.~\eqref{eq:seqEcho} and \eqref{eq:seqDD}. Results are shown in Fig.~\ref{fig:dd}(b) for $\omega_q \Delta t= 0.1$.
Apparently, in the very-short-time region after the application of a single Hahn-pulse a characteristic growth of coherences can be seen.
In the short-time regime $\omega_c \Delta t' \ll 1$, the fluctuation of $\hat{V}$ caused by the reservoir can be seen as static, and thus the recovery of coherences is observed irrespective of the spectral exponent $s$.
Right after this \emph{universal recovery} of coherences, the system experiences decoherence when the duration of the first idle phase $\Delta t$ is small. 
This decoherence is similar to the universal decoherence \cite{TuorilaPRR2019,BraunPRL2001}:
Due to the relation $[\hat{H}_S, \hat{V}] = 0$, the decay of the coherence is determined only by the properties of the reservoir, whose representation is the same as the one for the universal decoherence (see Appendix~\ref{sec:appEcho}).
The universal decoherence is slower for deeper sub-Ohmic reservoirs, which in turn leads to larger timescales for the universal recovery.
By contrast, when $\Delta t$ increases, further recovery of coherence is observed for deep sub-Ohmic reservoirs, cf.\ inset of Fig.~\ref{fig:dd}(b) for $\omega_q \Delta t =1$ with $s = 1/14$: a fast increase of coherence in the region $1 \leq \omega_q (\Delta t + \Delta t') \leq 1.05$ is found and a subsequent slower increase, rather than universal decoherence, in the region $1.05 \leq \omega_q (\Delta t + \Delta t')$.
We attribute the former steep increase to the universal behavior and the bending towards a slower increase to the recovery found also in HE (we refer to this recovery as an HE-like recovery hereafter).

We conclude from this analysis that, if for a DD sequence, 
the next pulse is applied within this short-time range, decoherence is indeed suppressed very efficiently. If the time between the first and the second pulse exceeds this time range, decoherence becomes substantial, very similar to what happens after application of a single HE.
\textcolor{black}{We here emphasize that this time range for the universal recovery is restricted by the cutoff frequency $\omega_c$, which implies the dependence of $\Delta t$ for the local minima on the cutoff frequency $\omega_c$ discussed above.}

Now, we turn to the analysis of the large-$\Delta t$ region, $\omega_q \Delta t \geq 0.2$.
In the Ohmic case, steep drops in $|\rho_{eg}(t)|$ are observed in Fig.~\ref{fig:dd}(c).
This is due to the relatively large portion of high-frequency modes in $S_\beta(\omega)$ as discussed above for the HE experiments. Larger values of $\Delta t$ lead to decreasing relevance of these modes and, consequently, the decay of coherence slows down as $\Delta t$ increases [left panel of Fig.~\ref{fig:dd}(c), $s = 1$]; it even reverts to an increasing time constant $T_\mathrm{DD}$
in the region $\omega_q \Delta t \geq 0.2$.

In the deep sub-Ohmic domain, the steep drops discussed above are no longer observed. Instead, the HE-like recovery of the intensity is observed in Fig.~\ref{fig:dd}(c) leading to an improved coherence time even beyond what can be achieved with HE and what is observed in a Ramsey experiment (no pulses). Here, the strong presence of low-frequency modes induces a smoother transition from the short- to the moderate- and the long-time range.
At $\omega_q \Delta t \approx 1$, the transition from the universal suppression of decoherence to the HE-like recovery appears. With increasing duration $\Delta t$ beyond the local minimum in $T_\mathrm{DD}$, decoherence between two successive pulses plays a significant role [right panel of Fig.~\ref{fig:dd}(c)], and the time constant decreases. For example, in the case $s = 1/14$, this results even in a local maximum at $\omega_q \Delta t = 3$.
The same discussion holds for the case $s = 1/8$.

The cases with $s = 1/2$ and $1/4$ are intermediate between those with $s = 1$ and $1/14$: In Fig.~\ref{fig:dd}(c) with $s = 1/2$, the red curve displays the steep drop, although the magnitude of this decay is smaller in comparison to the case $s = 1$ [see the inset of Fig.~\ref{fig:dd}(c) for $s = 1/2$].
By contrast, the blue curve exhibits slight recoveries of the intensity in the middle and at the end of each cycle. This indicates that in the small-$\Delta t$ region, $0.2 \leq \omega_q \Delta t \lesssim 4$, an improvement related to the fast decay is observed while in the longer $\Delta t$ region, $\omega_q\Delta t \gtrsim 5$, the HE-like recovery appears.

We thus draw the following general conclusions. (i) The time constant $T_\mathrm{DD}$ strongly depends on the duration $\Delta t$ and the spectral exponent $s$.
We would like to emphasize that the performance depends on the detailed profile of the spectral noise power: the comparison between $\omega_c$ and $\Delta t$, which has been conducted in previous studies~\cite{Lidar2013}, might be insufficient in some cases.
(ii) For Ohmic reservoirs, the DD approach never improves the time constant compared to HE and Ramsey experiments, unless for extremely short times $\omega_c \Delta t < 1$ with $\omega_q\ll \omega_c$.
This corresponds to the ``decoherence acceleration'' found also in previous studies~\cite{ViolaPRA1998,RegoARPC2009}.
(iii) For sub-Ohmic spectral densities with $1/4< s\leq 1$ the application of DD is beneficial compared to Ramsey experiments but the time constant still remains below the one seen in HE except for very short times.
(iv) In the relatively deep sub-Ohmic range $s\leq 1/4$, the DD provides an improvement of the stability of coherence beyond both the HE and Ramsey findings for almost all parameter values; a few exceptions are found in our results for $\omega_q \Delta t = 0.2$ and $1$ with $s = 1/4$.


Now, we comment on predictions in previous studies.
The prediction of the decrease in the coherence time $T_\mathrm{DD}$ in the very-short-time region $\omega_c \Delta t < 1$ in Refs.~\cite{ViolaPRL1999,KhodjastehPRA2007} is consistent with our results.
In previous experiments~\cite{AlvarezPRL2011,SouzaPRL2011}, local maxima of $T_\mathrm{DD}$ were found, which was attributed to imperfect pulses.
Our results suggest that local maxima are observed even with perfect pulses, for timescales sufficiently larger than those of the cutoff frequency of the reservoir.

As a final technical remark, we mention that our simulations provide somewhat less accurate results in the very-short-time region $\omega_q \Delta t \leq 0.02$.
As illustrated in Fig.~\ref{fig:dd}(e), the decay is very slow in the region $\omega_q \Delta t \leq 0.02$ (orders of magnitudes larger $T_\mathrm{DD}$ than that for longer $\Delta t$) so that numerical calculations are extremely susceptible to even tiny numerical errors. We observed, however, that the degrading accuracy of the results does not change the qualitative profile in Fig.~\ref{fig:dd}(e), that is, the decrease of $T_\mathrm{DD}$ up to the duration $\omega_q \Delta t = 0.2$ as well as the acceleration of decoherence for $\omega_q \Delta t \geq 0.02$ in the Ohmic case.


\section{Concluding remarks} \label{sec:conclusion}
In this paper, we numerically studied pure-dephasing dynamics of a single qubit in presence of realistic time-retarded noise, focusing on effects of the static qubit--reservoir correlation at an initial time and the performance of typical quantum control techniques, such as Hahn echo (HE) and dynamical decoupling (DD).
\textcolor{black}{For further understanding of the qubit's dynamics and the applicability of those control techniques, we chose specific spectral densities relevant for various qubit platforms, in particular for superconducting quantum devices.}
The class of the spectral densities considered in this study ranges from typical Ohmic noise, which corresponds to white noise at higher temperatures, to deep sub-Ohmic noise, which represents $1/f^\varepsilon$-type noise in a small-frequency region.
For the DD simulations, we studied the whole dynamics of a single qubit in time with a varying time interval between two subsequent pulses (duration of the idle phase).
This contrasts with previous studies, which focused on a stroboscopic behavior at the end of a cycle in the small interval limit $\Delta t \to 0$.

On the basis of new analytical expressions, we analyzed here the most common schemes to fight against dephasing noise.
For the Ramsey experiments we found a shift of the effective Larmor frequency caused by the correlated equilibrium state $e^{-\beta\hat{H}_\mathrm{tot}}$.
The deviation from the frequency of the bare system $\omega_q$ depends on the profile of the spectral density:
the slow decay of the two-time correlation function, i.e., the large portion of low-frequency modes of the spectral density, contributes to the frequency shift.
It is worth noting that not only the Lamb shift, which originates from the correlation between the system and reservoir during the real-time evolution, but also the initial correlation affects the effective Larmor frequency.

For the HE sequences, the improvement of the coherence time was quantitatively analyzed.
Exploring the whole dynamics during the HE sequence, we found that the recovery of the coherence $|\rho_{eg}(t)|$ occurs in the sub-Ohmic case ($s \leq 1/2$), while the echo pulse always deteriorates the coherence in the Ohmic case ($s = 1$).
As theoretically predicted, the improvement is most significant for the deepest sub-Ohmic bath ($s=1/14$), which generates the \emph{nearly static} noise.

In the DD simulation, the performance of the CPMG sequence was investigated. 
As expected, we found that when the timescale of the pulse interval is sufficiently shorter than those for the reservoir cutoff frequencies ($\omega_c \Delta t \ll 1$), coherences are stabilized irrespective of the bath spectral density, which corresponds to previous studies~\cite{ViolaPRA1998,Lidar2013}.
The main focus has been on the performance for longer pulse durations in agreement with the experimental situation:
Interestingly, for sub-Ohmic baths coherence times are substantially enhanced compared to simple Ramsey results due to reservoir-induced feedback.
For reservoirs with exponents $s < 1/2$, even local maxima of coherence times can be identified around pulse intervals on the order of the qubit transition frequency.
By contrast, the coherence time for DD is worse than that for Ramsey experiments in the Ohmic case ($s=1$) once the condition $\omega_c \Delta t \ll 1$ is violated.
We conclude that neither HE nor DD can stabilize coherence in the pure Ohmic case.
Overall, we have obtained a comprehensive picture of the DD performance depending on not only the time interval $\Delta t$ but also the profile of the spectral noise power, especially the spectral exponent $s$.

\textcolor{black}{The main general findings of this study corresponding to the questions raised in Sec.~\ref{sec:Introduction} are thus (proceeding in reverse order):}
\textcolor{black}{(Q2) What is the impact of initial correlations between qubits and environments?:
The initial correlations induced by the total equilibrium state cause the shift of the Larmor frequency when the system is subject to the pure-dephasing noise.
Since the correlations only shift the frequency, they hardly affect the performance of the HE and DD in terms of the absolute amplitude of the coherence, $|\rho_{eg}(t)|$, and coherence time,  $T_\mathrm{E}$ and $T_\mathrm{DD}$.}

\textcolor{black}{(Q1) What is the performance of DD schemes beyond the short-time assumption for various noise sources?:
The performance with moderate time intervals ($\omega_q \Delta t \gtrsim 1$) highly depends on the form of the noise spectral density, i.e., the spectral exponent.
For Ohmic noise,  longer idle times $\Delta t$  between subsequent pulses improve the performance, while there exists an optimal $\Delta t$  for deep sub-Ohmic reservoirs.}

\textcolor{black}{To go beyond this conceptual insight and to provide explicit predictions for actual experimental settings, however, the modeling must be extended. This includes at least two main features, namely, to generalize the coupling form between the system and reservoir beyond the pure-dephasing type and to lift the assumption of  ideal impulsive pulses with zero width.}
\textcolor{black} {To go beyond the pure dephasing model, one has to include population-relaxation processes ($\hat{V} = \hbar\hat{\sigma}_x$), which are known to affect the coherence time of  Ramsey experiments~\cite{Breuer2002}.
To answer to what extent population-relaxation processes are detrimental to the CPMG sequence~\cite{Lidar2013}, one can employ the  methods as in Refs.~\cite{Nakamura18PRA,tanimuraJCP20,XuPRL2022} to treat this general situation in a numerically rigorous manner.}
\textcolor{black}{To express imperfect impulsive pulses, the obvious option is to change the axis and angle of the rotation operator in Eq.~\eqref{eq:piPulse} from $\hat{R}_x(\pi) = \exp[-i \pi \hat{\sigma}_x/2]$ to, for example, $\hat{R}_\mathrm{err}(\pi; \delta, \varepsilon) = \exp[-i (\pi + \delta)(\hat{\sigma}_x \cos \varepsilon + \hat{\sigma}_y \sin \varepsilon)/2]$ ($\delta$ and $\varepsilon$ are small errors).
With this change, the diagonal elements of the RDO also contribute to the dynamics of the off-diagonal elements.
Finite-width pulses can be implemented in our method by replacing $\hat{R}_x(\pi)$ in Eq.~\eqref{eq:piPulse} with a corresponding time-evolution operator, $\hat{R}_x(\pi) \to \exp[-i\int dt' \hat{H}_\mathrm{tot}(t')/\hbar]$, where the time-dependent total Hamiltonian $\hat{H}_\mathrm{tot}(t)$ includes both dissipative couplings to reservoirs and time-dependent external fields corresponding to the pulse application.
It is worth noting that when the pulse duration is finite, the dynamics during the pulse application also affects the following idle period because of the time-retarded feedback of low-temperature reservoirs as shown explicitly for sequences of gate operations in Ref.~\cite{NakamuraPRR2024}. We will apply this methodology in a future work to explore these feedback effects, which in turn may necessitate optimized and even more sophisticated DD protocols.}

\section*{Acknowledgement}
The authors would like to thank J.~T.~Stockburger for fruitful discussions. This work was supported by the BMBF through QSolid and the Cluster4Future QSens (Project QComp) and the DFG through Grant No. AN336/17-1 (FOR2724).
K.N. acknowledges support from the State of Baden-W{\"u}rttemberg through bwHPC (Justus II).

\appendix
\section{The path-integral representation of the reduced density operator for pure-dephasing simulations} \label{sec:appEcho}
In this section, we derive equations for the dynamics of the reduced density operator (RDO) controlled by the Hahn-echo and dynamical-decoupling schemes, which was discussed in Sec.~\ref{sec:results}.
In those simulations, we only considered the impulsive pulses, and therefore we only need to consider the time evolution of the idle phases.
For those simulations, a quantum Gaussian environment is considered, leading to the Hamiltonian in the form of
\begin{align}
    \hat{H}_R = \sum_j \biggl(\frac{\hat{p}^2_j}{2m_j} + 
    \frac{1}{2} m_j \omega_j^2 \hat{x}^2_j\biggr)\, , 
    && 
    \hat{X} = \sum_j c_j \hat{x}_j
\end{align}
in Eq.~\eqref{eq:H_tot}.
Namely, the reservoir consists of an infinite number of harmonic oscillators, with $\hat{p}_j, \hat{x}_j, m_j$, and $\omega_j$ being the momentum, position, mass, and angular frequency of the $j$th bath, respectively.
The coupling strength between the system and $j$th oscillator, $c_j$, defines the spectral density as
\begin{align}
    J(\omega) = \sum_j \frac{c^2_j}{2m_j\omega_j} \delta(\omega - \omega_j)\, .
\end{align}
When we introduce the eigenvectors of the system Hamiltonian, $\hat{\sigma}_z \ket{a} = (-1)^{a+1}\ket{a}$ ($a = 0, 1$), and the eigenvectors of the position operator of the reservoir, $\ket{\bm{x}} = \ket{x_1, \ldots, x_j, \ldots}$, the time-evolution operator for the total Hamiltonian in Eq.~\eqref{eq:H_tot} is evaluated as 
\begin{align}
    & \Bigl\langle b, \bm{x}' \Big| e^{-i\hat{H}_\mathrm{tot}t/\hbar} \Big|a, \bm{x} \Bigr\rangle \\
    = & \delta_{ab} e^{-i\omega_a t}
    \biggl\langle\bm{x}' \bigg| \exp\biggl[-\frac{i}{\hbar}\Bigl(\hat{H}_R
    - (-1)^{a+1} \hbar\hat{X} \Bigr)t\biggr]\bigg|\bm{x}\biggr\rangle\, ,
\end{align}
where $\delta_{ab}$ is the Kronecker delta and $\omega_a = (-1)^{a+1} \omega_q / 2$.
The Boltzmann distribution $e^{-\beta\hat{H}_\mathrm{tot}}$ is also evaluated in the same way with the replacement of $i t / \hbar$ with $\beta$.
Note that the system part of the total Hamiltonian is replaced with $c$ numbers, and the bracket only includes the reservoir operators.
Considering the time evolution, $e^{-i\hat{H}_\mathrm{tot}t/\hbar} \hat{\rho}^\mathrm{c}_\mathrm{tot}(0) e^{i\hat{H}_\mathrm{tot}t/\hbar}$, where $\hat{\rho}^\mathrm{c}_\mathrm{tot}(0)$ is given by Eq.~\eqref{eq:initEchoC}, we obtain the matrix element of the RDO $\rho_{ab}(t) = \mel{a}{\hat{\rho}_S(t)}{b}$ in the following form by tracing out the reservoir degrees of freedom:
\begin{align}
    &\rho_{ab}(t) \\ =
    &e^{-i\omega_{ab}t} \sum_{c} \Bigl\langle a \Big| \hat{R}_y\Bigl(-\frac{\pi}{2}\Bigr) \Big| c \Bigr\rangle
    \frac{e^{-\beta \hbar \omega_c}}{Z} \Bigl \langle c \Big| \hat{R}_y\Bigl(\frac{\pi}{2}\Bigr) \Big| b\Bigr\rangle\\
    &\times \exp\biggl[
    \begin{aligned}[t]
        &-\int_{0}^{t} dt'\int_{0}^{t'} dt'' v_{ab}^2 C'(t'-t'') \\
        &+i \int_{0}^{t} dt' \int_{0}^{\beta\hbar}d\tau'
        v_{ab} \bar{C}(-t'-i\tau')(-1)^{c+1}\biggr]\, ,
    \end{aligned} \\
    \label{eq:FID1}
\end{align}
where $\omega_{ab} = \omega_{a} - \omega_{b}$ is the difference of the frequency between the bra and ket vectors, and $Z = \mathrm{tr}\{e^{-\beta \hat{H}_S}\} = 2 \cosh(\beta \hbar \omega_q /2)$ is the partition function of the bare system.
Note that the first term of the exponent corresponds to the conventional influence functional introduced in the Feynman--Vernon path integral representation~\cite{Feynman2017}, and $v_{ab} = (-1)^{a+1}-(-1)^{b+1}$ corresponds to the commutator $[\hat{V}, \bullet]$.
The contribution of the imaginary part of the two-time correlation function $C''(t)$ is always zero in our study.

The second term of the exponent originates from the correlated initial state and describes correlations between the Boltzmann distribution of the total Hamiltonian $e^{-\beta \hat{H}_\mathrm{tot}}$ and the time-evolution operator $e^{\pm i \hat{H}_\mathrm{tot}t/\hbar}$.
The extended two-time correlation function is defined as
\begin{align}
    \bar{C}(z) = \hbar\int_{0}^{\infty} d\omega J(\omega)
    \frac{\cosh\Bigl(\frac{\beta \hbar \omega}{2} - i \omega z\Bigr)}{\sinh\Bigl(\frac{\beta \hbar\omega}{2}\Bigr)}\, .
\end{align}
If the variable $z$ is a real number, $z \in \mathbb{R}$, the function $\bar{C}(z)$ coincides with $C(t)$.

Note that the influence functional with the correlated initial state includes the term
\begin{align}
    \exp\Biggl[\int_{0}^{\beta \hbar} \hspace*{-1ex} d\tau'
    \int_{0}^{\tau'} \hspace*{-1ex}d\tau''
    \bar{C}(-i\tau'+i\tau'')\Biggr] = e^{-\beta \lambda}\, ,
    \label{eq:reorg}
\end{align}
which is derived under the condition $[\hat{H}_S, \hat{V}] = 0$.
The quantity $\lambda = \hbar^{2} \int_{0}^{\infty} d\omega J(\omega) / \omega$ is the reorganization energy.
This term only shifts the origin of the energy and is omitted with the normalization condition at $t = 0$.

For the Hahn-echo experiment, the pulse sequence in the superoperator representation is given by Eq.~\eqref{eq:seqEcho}, and corresponding time evolution is evaluated as
\begin{align}
    & \Bigl\langle b, \bm{x}'\Big| e^{-i\hat{H}_\mathrm{tot}\Delta t'/\hbar}
    \hat{R}_x(\pi)e^{-i\hat{H}_\mathrm{tot}\Delta t/\hbar} \Big| a, \bm{x} \Bigr\rangle \\
    = & e^{-i(\omega_b \Delta t' + \omega_a \Delta t)}
    \mel{b}{\hat{R}_x(\pi)}{a} \\
    & \times \biggl\langle \bm{x}' \biggr|
    \exp\biggl[-\frac{i}{\hbar} \Bigl(\hat{H}_R
    - (-1)^{b+1}\hbar\hat{X}\Bigr)\Delta t'\biggr] \\
    & \times \exp\biggl[-\frac{i}{\hbar} \Bigl(\hat{H}_R
    - (-1)^{a+1}\hbar\hat{X}\Bigr)\Delta t\biggr]
    \biggl| \bm{x} \biggr\rangle \, .
\end{align}
By using this relation, the matrix element of the RDO at the time $t_\mathrm{e} = \Delta t' + \Delta t$ is derived as
\begin{align}
    & \rho_{ab}(t_\mathrm{e}) \\
    = & \sum_{c, d, e}\;
    e^{-i(\omega_{ab}\Delta t' 
    + \omega_{de} \Delta t)} \mel{a}{\hat{R}_x(\pi)}{d} \hspace*{-0.5ex}\mel{e}{\hat{R}_x(-\pi)}{b} \\
    & \times \Bigl\langle d \Big| \hat{R}_y\Bigl(-\frac{\pi}{2}\Bigr) \Big| c \Bigr\rangle
    \frac{e^{-\beta\hbar\omega_c}}{Z}
    \Bigl\langle c \Big| \hat{R}_y\Bigl(\frac{\pi}{2}\Bigr) \Big| e \Bigr\rangle
    F(t_\mathrm{e})\, , \\
    \label{eq:echo1}
\end{align}
where
\begin{align}
    F(t_\mathrm{e}) = 
    & \exp\biggl[
    \begin{aligned}[t]
        & -\int_{0}^{t_\mathrm{e}} \hspace*{-2ex} dt' \int_{0}^{t'} \hspace*{-2ex} dt''
        v^\times(t') C'(t'-t'')v^\times(t'') \\
        & +i \int_{0}^{t_\mathrm{e}} \hspace*{-2ex} dt' \int_{0}^{\beta\hbar} \hspace*{-2.5ex} d\tau'
        v^\times(t') \bar{C}(-t'-i\tau')(-1)^{c+1}\biggr]\, .
    \end{aligned}
    \label{eq:IF4}
\end{align}
Here, $v^\times(t) = v(t) - v'(t)$ is the commutator, and we have defined $v(t)$ and $v'(t)$ as
\begin{align}
    v(t) & = \Biggl\{
    \begin{aligned}
        & (-1)^{d+1} && (0 \leq t \leq \Delta t) \\
        & (-1)^{a+1} && (\Delta t \leq t \leq \Delta t + \Delta t')
    \end{aligned} \, , \\
    v'(t) & = \Biggl\{
    \begin{aligned}
        & (-1)^{e+1} && (0 \leq t \leq \Delta t) \\
        & (-1)^{b+1} && (\Delta t \leq t \leq \Delta t + \Delta t')
    \end{aligned} \, .
\end{align}
Again, the imaginary part $C''(t)$ does not contribute to the dynamics.

For the dynamical-decoupling simulations, the time-evolution operator corresponding to Eq.~\eqref{eq:seqDD} is expressed as
\begin{align}
    & e^{-i\hat{H}_\mathrm{tot}t_{2n+1}/\hbar} \hat{R}_x(\pi)
    e^{-i\hat{H}_\mathrm{tot}t_{2n}/\hbar} \hat{R}_x(\pi)\\
    & \times \cdots \times
    \hat{R}_x(\pi)e^{-i\hat{H}_\mathrm{tot}t_{2}/\hbar} \hat{R}_x(\pi)
    e^{-i\hat{H}_\mathrm{tot}t_{1}/\hbar}\, ,
\end{align}
where $t_{1} = \Delta t /2$ and $t_2 = t_3 = \cdots = t_{2n} = \Delta t$. At the end of the CPMG sequence, the value $t_{2n+1}$ is given by $\Delta t/2$, while we vary $t_{2n+1}$ to numerically obtain the whole dynamics. The matrix element of the RDO $\rho_{a_{2n+1} b_{2n+1}}(t_\mathrm{e})$ at the time $t_\mathrm{e} = \sum_{l=1}^{2n+1} t_l$ is described as
\begin{align}
    & \hat{\rho}_{a_{2n+1} b_{2n+1}} (t_\mathrm{e}) \\
    = & e^{-i \omega_{a_{2n+1} b_{2n+1}}t_{2n+1}}
    \!\!\sum_{\substack{a_1, \ldots, a_{2n} \\ b_1, \ldots, b_{2n}}} \!\!
    F(t_\mathrm{e})\\
    & \times \sum_{c} \Bigl\langle a_1 \Big| \hat{R}_y\Bigl(-\frac{\pi}{2}\Bigr) \Big| c \Bigr\rangle
    \frac{e^{-\beta\hbar\omega_c}}{Z} \Bigl\langle c \Big| \hat{R}_y\Bigl(\frac{\pi}{2}\Bigr) \Big| b_1 \Bigr\rangle \\
    & \times \prod_{l=1}^{2n} e^{-i\omega_{a_l b_l}t_l}
    \mel{a_{l+1}}{\hat{R}_x(\pi)}{a_l} \hspace*{-0.5ex} \mel{b_l}{\hat{R}_x(-\pi)}{b_{l+1}}\, .
    \\
    \label{eq:DD1}
\end{align}
The function $F(t_\mathrm{e})$ takes the same form as Eq.~\eqref{eq:IF4}, but $v(t)$ and $v'(t)$ take different forms as follows:
\begin{align}
    v(t) & = (-1)^{a_{l}+1}\, , \\
    v'(t) & = (-1)^{b_{l}+1}\, ,
\end{align}
which are defined for the interval $\tilde{t}_{l-1} \leq t \leq \tilde{t}_l$ ($l=1, \ldots, 2n+1$), where $\tilde{t}_0 = 0$ and $\tilde{t}_l = \sum_{j=1}^{l} t_j$ $(l \geq 1)$. 
To evaluate the density operator with the factorized initial state, we replace the initial Boltzmann distribution in Eqs.~\eqref{eq:FID1}, \eqref{eq:echo1}, and \eqref{eq:DD1} as $e^{-\beta\hbar\omega_0}/Z \to 1$ and $e^{-\beta\hbar\omega_1}/Z \to 0$ and set the term $(-1)^{c+1}$ in the influence functional to $0$.

Here, we describe the concrete form of the off-diagonal element of the RDO with the factorized initial state, utilizing the  relation $\rho_{eg}(t) = \rho_{10}(t)$.
For the Ramsey experiment, in which no pulses are applied, it is derived as 
\begin{align}
    & \rho_{eg}(t) \\
    = & \frac{1}{2} e^{-i\omega_q t}
    \exp\Biggl[-4\int_{0}^{t}dt'\int_{0}^{t'}dt'' C'(t'-t'') \Biggr] \\
    = & \frac{1}{2} e^{-i\omega_q t} 
    \exp\Biggl[-4\hbar\int_{0}^{\infty} d\omega J(\omega)
    \coth\frac{\beta\hbar\omega}{2}
    \frac{1-\cos \omega t}{\omega^2}\Biggr]\, . \\
    \label{eq:FID2}
\end{align}
For the echo experiment with $\Delta t' = \Delta t$, we obtain
\begin{align}
    & \rho_{eg}(t_\mathrm{e} = 2\Delta t) \\
    = & \frac{1}{2} \exp\biggl[-4\hbar\int_{0}^{\infty} d\omega
    \begin{aligned}[t]
        & J(\omega) \coth \frac{\beta\hbar\omega}{2} \\ 
        & \times \frac{1-\cos 2\omega \Delta t}{\omega^2} \tan^2
        \frac{\omega\Delta t}{2}\biggr]\, ,
    \end{aligned} \\
    \label{eq:echo2}
    \noeqref{eq:echo2}
\end{align}
and for the dynamical decoupling with $t_{2n+1} = 0$ and $t_{1} = \Delta t$ (an asymmetric version of the CPMG sequence), we obtain
\begin{align}
    & \rho_{eg}(t_\mathrm{e} = 2n\Delta t) \\
    = & \frac{1}{2} \exp\biggl[-4\hbar\int_{0}^{\infty} d\omega
    \begin{aligned}[t]
        & J(\omega) \coth \frac{\beta\hbar\omega}{2} \\ 
        & \times \frac{1-\cos 2n\omega \Delta t}{\omega^2} \tan^2
        \frac{\omega\Delta t}{2}\biggr]\, .
    \end{aligned}\\
    \label{eq:DD2}
\end{align}
Note that Eqs.~\eqref{eq:FID2}--\eqref{eq:DD2} correspond to equations in previous studies~\cite{IthierPRB2005,ViolaPRA1998,ShiokawaPRA2004,RegoARPC2009}.
It is worth noting that the form of Eq.~\eqref{eq:FID2} corresponds to that for the universal decoherence of the diagonal element of the RDO~\cite{TuorilaPRR2019,BraunPRL2001, NakamuraPRR2024}.

\subsection*{Dynamics with Markov approximation}
By differentiating Eq.~\eqref{eq:FID2}, we obtain the equation of motion for the off-diagonal element of the RDO as follows:
\begin{align}
    \dot{\rho}_{eg}(t) = -\biggl[i \omega_q + 4 \int_0^{t} dt'' C'(t-t'')\biggr] \rho_{eg}(t)\, .
\end{align}
If we impose the Markov approximation, in which the two-time correlation function is localized at the time $t \simeq 0$, we can change the lower limit of the integration from $0$ to $-\infty$, and the equation reads
\begin{align}
    \dot{\rho}_{eg}(t) = -(i \omega_q + \gamma_\mathrm{pd}) \rho_{eg}(t)\, .
\end{align}
The pure-dephasing rate $\gamma_\mathrm{pd}$ is expressed as 
\begin{align}
    \gamma_\mathrm{pd} = 4\pi \lim_{\omega \to 0} S_\beta(\omega) 
    = 2 \pi \hbar \lim_{\omega \to 0} J(\omega) \coth \frac{\beta \hbar \omega}{2}\, ,
    \label{eq:gammaPd}
\end{align}
which coincides with the rate obtained from the Lindblad equation~\cite{Breuer2002} and noninteracting-blip approximation (NIBA)~\cite{Weiss2012}.
As discussed in the main text, the above value diverges in the sub-Ohmic case, and this argument cannot be applied.

\textcolor{black}{Solving this equation, we obtain the dynamics of the off-diagonal element within the Markov approximation in the form of $\rho_{eg}(t) = 0.5e^{-(i\omega_q + \gamma_\mathrm{pd})t}$, with the initial condition $\rho_{eg}(0) = 0.5$.
For the Ohmic case, the time constant is evaluated as $T_\mathrm{R} = 1/\gamma_\mathrm{pd} = \beta / (4 \pi \kappa) = 62.5 / \omega_q$, which is in agreement with the exact results in Table~\ref{tbl:timeConst}.
Despite the agreement of the time constant, however, the whole dynamics during the Ramsey sequence are different, as discussed in the main text [Fig.~\ref{fig:echo3d}(a), $s=1$].}

\textcolor{black}{For the HE experiment, the time derivative of $\rho_{eg}(\Delta t + \Delta t')$ with respect to $\Delta t'$ is derived as [cf.\ Eqs.~\eqref{eq:echo1} and \eqref{eq:IF4}]
\begin{align}
    & \frac{\partial \rho_{eg}(\Delta t + \Delta t')}{\partial \Delta t'} \\ = &-
    \begin{aligned}[t]
      \biggl\{ & i\omega_q 
      + 4 \int_{\Delta t}^{\Delta t + \Delta t'} \hspace{-6ex} dt'' C'(\Delta t + \Delta t' - t'') \\
      & - 4 \int_{0}^{\Delta t} \hspace{-2ex} dt'' C'(\Delta t + \Delta t ' - t'')\biggr\}
      \rho_{eg}(\Delta t + \Delta t')
    \end{aligned}
\end{align}
with the factorized initial states.
When the two-time correlation function is well approximated as $C'(t) \simeq \delta (t)$, the third term can be neglected, and the second term is again approximated with $\gamma_\mathrm{pd}$.
Hence, the decoherence dynamics during the HE sequence are evaluated as $|\rho_{eg}(t)| = 0.5 e^{-\gamma_\mathrm{pd}t}$ and coincide with those obtained with the Ramsey experiments within the Markov approximation.
The same argument holds for the DD case, and it is concluded that the non-Markovianity plays a crucial role in the recovery of the coherence.}

\textcolor{black}{For the simulation of the HE experiments with the Markov approximation in the case $s = 1/14$ in \textcolor{black}{Fig.~\ref{fig:echo3d}(e)}, we exploited the extracted value $T_\mathrm{R}$ in Table~\ref{tbl:timeConst} for $\gamma_\mathrm{pd}$, since we cannot use Eq.~\eqref{eq:gammaPd} in the sub-Ohmic case, as mentioned above.}

\subsection*{Dynamical decoupling in small-$\Delta t$ limit}
Due to the cutoff function, $J(\omega)$ fast decays to $0$ in the region $\omega \geq \omega_c$.
When we consider the condition $\omega_c \Delta t \ll 1,$ which corresponds to the small $\Delta t$-limit, the term $\tan^2(\omega \Delta t /2)$ in the integrand in Eq.~\eqref{eq:DD2} is approximated with $(\omega \Delta t/2)^2$. If the number $n$ is large enough, the term $\cos 2n \omega \Delta t$ exhibits fast oscillation with respect to $\omega$, and its contribution to the integral becomes negligibly small. Consequently, the off-diagonal element at a long enough time is evaluated as 
\begin{align}
    & \rho_{eg}(t_\mathrm{e}=2n\Delta t) \\
    \simeq & \frac{1}{2} \exp\biggl[-\hbar \Delta t^2 
    \int_{0}^{\infty} d\omega J(\omega) 
    \coth \frac{\beta \hbar \omega}{2}
    \biggr]\, ,
    \label{eq:DD2Asympt}
\end{align}
which corresponds to asymptotic saturation pointed out in a previous study~\cite{RegoARPC2009}.
We confirmed that the exponent of Eq.~\eqref{eq:DD2Asympt} is smaller for the smaller spectral exponent $s$ (deeper sub-Ohmic reservoirs) in our simulation.
This indicates that the saturated value of $\rho_{eg} (t_\mathrm{e} = 2 n \Delta t \gg 1/\omega_q)$ is larger for the smaller $s$ in the small-$\Delta t$ limit.
Assuming that the function $\rho_{eg}(2n\Delta t)$ monotonically decays with respect to $n$, which is true in our study, we deduce that the time constant $T_\mathrm{DD}$ is larger when the spectral exponent $s$ decreases.
This prediction deviates from the numerical results in Fig.~\ref{fig:dd}(e), which results from the numerical errors discussed in the main text.
Note that Eq.~\eqref{eq:DD2Asympt} is derived on the basis of an asymmetric version of the CPMG sequence ($\Delta t_{2n+1} =0$ and $t_1 = \Delta t$), while the sequence in the main text is a symmetric version ($\Delta t_{2n+1} = \Delta t_{1} = \Delta t /2$).

\subsection*{Numerical implementation}
To obtain numerical results, we evaluate the time derivatives, $\partial \rho_{eg}(t) / \partial t$ for Eq.~\eqref{eq:FID1} and $\partial \rho_{eg}(t_\mathrm{e}) / \partial t_\mathrm{e}$ for Eqs.~\eqref{eq:echo1} and \eqref{eq:DD1}.
For both factorized and correlated initial states, we need to evaluate the function
\begin{align}
    \gamma'(t;t_2,t_1) = \int_{t_1}^{t_2} dt'' C'(t-t'')\, ,
    \label{eq:intCorr}
\end{align}
which appears in those time derivatives.
Note that we can compute this function with the aid of the representation of the two-time correlation function in Eq.~\eqref{eq:CF}.

The function $\bar{L}(t)$ in Eq.~\eqref{eq:FIDC} is the integral of $\bar{C}(-t'-i\tau')$ with respect to $\tau'$ and given by
\begin{align}
    \bar{L}(t) = & \hbar \int_{0}^{\infty} d\omega \frac{J(\omega)}{\omega} \cos\omega t \\
    = & \frac{1}{2} \int_{0}^{\beta\hbar} d\tau' \bar{C}(-t-i\tau')
    = -\int_{t}^{\infty} dt' C''(t')\, .
\end{align}
Note that the relaxation function of the reservoir $\Psi(t)$ is proportional to $\bar{L}(t)$, which is expressed as $\Psi(t) = 2 \bar{L}(t) / \hbar$~\cite{Kubo1985}.
For the Hahn echo and dynamical decoupling, the off-diagonal element is given by $\rho_{eg}(t) = \rho_{+}(t) - \rho_{-}(t)$ with the following replacement in Eq.~\eqref{eq:FIDC}:
\begin{align}
    \omega_q t 
    & \to \omega_q (\Delta t' - \Delta t) && \mbox{(Hahn echo)}\, , \\
    & \to \omega_q \sum_{l=1}^{2n+1} (-1)^{l+1} t_l  && \mbox{(dynamical decoupling)}\, ,
\end{align}
\begin{align}
    4\int_{0}^{t}dt'\int_{0}^{t'} & dt'' C'(t'-t'') \\
    & \to \int_{0}^{t}dt'\int_{0}^{t'} dt'' v^\times(t')C'(t'-t'') v^\times(t'')\, ,
\end{align}
\begin{gather}
    4i\int_{0}^{t}dt'\bar{L}(t') \to 2i\int_{0}^{t}dt' v^\times(t') \bar{L}(t')\, .
\end{gather}
Similar to the evaluation of $\gamma'(t;t_2, t_1)$ in Eq.~\eqref{eq:intCorr}, the function $\bar{L}(t)$ is computed with the aid of Eq.~\eqref{eq:CF}:
we solve the equation $d \bar{L}(t) / dt = C''(t)$ with the initial condition $\bar{L}(0) = \lambda / \hbar$ [$\lambda$ is the reorganization energy in Eq.~\eqref{eq:reorg}].

\section{Time constants for the Ramsey, echo, and dynamical-decoupling experiments} \label{sec:appTimeConst}
In this appendix we explain the fitting methods to obtain the time constants in Sec.~\ref{sec:results}.
We conducted the method of least squares and evaluated the time constants $T_\mathrm{R}$, $T_\mathrm{E}$, and $T_\mathrm{DD}$.
In the Ramsey experiments for $s = 1$ and $1/2$, fast decay of $|\rho_{eg}(t)|$ was observed.
Therefore, we considered the following functions as the model curves:
\begin{align}
    |\bar{\rho}_{eg}(t)|_{(1)} &= 
    A \exp[-B t] + C \exp[-D t] +E\, ,
    \label{eq:twoExp} \\
    |\bar{\rho}_{eg}(t)|_{(2)} & =
    A \exp[-B t] + C \exp[- (D t)^2] + E\, , \quad
    \label{eq:Exp+Gauss}
\end{align}
and compared the residuals of the method of least squares.
In the case for $s = 1$, a smaller residual was obtained with the curve for Eq.~\eqref{eq:twoExp} compared to the curve for Eq.~\eqref{eq:Exp+Gauss}, while the opposite result was obtained in the case $s = 1/2$.
In both cases, $T_\mathrm{R}$ was evaluated with the relation $ T_\mathrm{R} = 1 / D$.
Here, we defined $D < B$ in the Ohmic case, and the relation $D < B$ holds in the case $s = 1/2$.

In the other cases of the Ramsey experiments, the cases of the HE experiments with the condition $s \neq 1$, and all the cases of the DD experiments, fast decay was not observed, and we considered an exponential function,
\begin{align}
    |\bar{\rho}_{eg}(t)|_{(3)} = a \exp[-b t] + c\, , 
\end{align}
and a Gaussian function, 
\begin{align}
    |\bar{\rho}_{eg}(t)|_{(4)} = a \exp[- (b t)^2 ] + c,
\end{align}
as the model curves. In the Ramsey cases for $s \leq 1/4$ and the HE cases for $s \neq 1$, we found that the Gaussian fitting was better than the exponential fitting.
For the DD experiments, we obtained better results with the exponential fitting compared to the Gaussian fitting.

For the HE experiment for $s = 1$, we evaluated the time constant from the asymptotic behavior of $|\rho_{eg}(t)|$:
we found from Fig.~\ref{fig:echo3d}(a) that the dynamics after the fast decay with different $\Delta t$ can be represented by a single curve. We used this curve for the evaluation of $T_\mathrm{E}$. The model curves $|\bar{\rho}_{eg}(t)|_{(3)}$ and $|\bar{\rho}_{eg}(t)|_{(4)}$ were considered, and the exponential fitting was found to be better. The time constants obtained with these curves are listed and plotted in Table~\ref{tbl:timeConst} and Figs.~\ref{fig:dd}(d) and \ref{fig:dd}(e), respectively.
\bibliography{reference,qubit,dd}
\end{document}